\documentclass[prc,preprint,showpacs,showkeys,lengthcheck,
  nofootinbib,tightenlines,onecolumn,notitlepage,preprintnumbers,superscriptaddress]{revtex4-1}

\usepackage[utf8]{inputenc}
\usepackage{newtxtext,newtxmath}
\usepackage[mathcal]{euscript}

\usepackage{graphicx}
\usepackage[usenames,dvipsnames]{xcolor}
\graphicspath{{figs/}}

\usepackage{hyperref}
\hypersetup{colorlinks=true,citecolor=blue,linkcolor=blue,urlcolor=blue}

\usepackage{diagbox}

\usepackage{amsmath,amssymb,amsfonts}
\usepackage{slashed}
\usepackage{bm}

\begin{document}
\title{The energy-momentum tensor of spin-1 hadrons: formalism}

\author{Wim Cosyn}
\email{wim.cosyn@ugent.be}
\affiliation{Department of Physics and Astronomy, Ghent University, 
             Proeftuinstraat 86, B9000 Ghent, Belgium}

\author{Sabrina Cotogno}
\email{sabrina.cotogno@polytechnique.edu}
\affiliation{Centre de Physique Th\'eorique, \'Ecole polytechnique, 
	CNRS, Universit\'e Paris-Saclay, F-91128 Palaiseau, France}

\author{Adam Freese}
\email{afreese@anl.gov}
\affiliation{Argonne National Laboratory, Lemont, Illinois 60439, USA}
	
\author{C\'edric Lorc\'e}
\email{cedric.lorce@polytechnique.edu}
\affiliation{Centre de Physique Th\'eorique, \'Ecole polytechnique, 
	CNRS, Universit\'e Paris-Saclay, F-91128 Palaiseau, France}

\begin{abstract}
    We provide the complete decomposition of the local gauge-invariant energy-momentum tensor for spin-1 hadrons, including non-conserved terms for the individual parton flavors and antisymmetric contributions originating from intrinsic spin.  We state sum rules for the gravitational form factors appearing in this decomposition and provide relations for the mass decomposition, work balance, total and orbital angular momentum, mass radius, and inertia tensor. Generalizing earlier work, we derive relations between the total and orbital angular momentum and the Mellin moments of twist-2 and 3 generalized parton distributions, accessible in hard exclusive processes with spin-1 targets.  Throughout the work, we comment on the unique features in these relations originating from the spin-1 nature of the hadron, being absent in the lower spin cases.
\end{abstract}

\date{\today}
\maketitle
\onecolumngrid

%%%%%%%% Introduction %%%%%%%%%%%%%%
\section{Introduction}
\label{sec:intro}

In recent years,
a lot of attention has been given to the energy-momentum tensor (EMT)
as a fundamental object of study in hadronic physics and QCD, see e.g.~\cite{Leader:2013jra,Polyakov:2018zvc} and references therein.
Hadronic matrix elements of the (local) EMT operator for quarks and gluons are
parametrized in terms of gravitational form factors (GFFs)\footnote{
  We call the form factors ``gravitational'' since the EMT is usually understood
  as the source of gravitational interactions.
  However, we do not measure in practice these form factors through gravity and
  we do not know the exact form that a theory of quantum gravity will take.
  Thus, it is unclear whether gravitation sees the symmetric, Belinfante-improved
  EMT as in general relativity or an asymmetric EMT as in Einstein-Cartan
  theory~\cite{Cartan:1923zea}.
  Despite this, we refer to all form factors appearing in either the symmetric
  or asymmetric EMT as ``gravitational'' form factors.
}, just as hadronic matrix elements of the charge current operator are parametrized in terms of electromagnetic form factors. The GFFs encode properties that are of great interest, such as the hadronic mass and angular momentum sum rules and their spatial distributions~\cite{Polyakov:2002yz,Leader:2013jra,Lorce:2017xzd,Lorce:2017wkb}, and can shed light on novel properties of hadrons, such as the way that stress and shear forces are distributed within them~\cite{Polyakov:2002wz,Polyakov:2002yz,Polyakov:2018zvc,Lorce:2018egm}. The topic places itself in the more general quest for a thorough understanding of the hadron structure, in which the spatial and momentum distributions of quarks and gluons play a crucial role.

Much of the recent literature on the hadron structure and QCD EMT has focused on either
spin-$1/2$~\cite{Lorce:2017xzd,Lorce:2017wkb,Polyakov:2018zvc,Polyakov:2018exb,Lorce:2018egm,Anikin:2019kwi}
or spin-$0$~\cite{Hudson:2017xug,Kumano:2017lhr,Tanaka:2018wea}
systems, see also~\cite{Alexandrou:2017oeh,Yang:2018bft,Yang:2018nqn,Shanahan:2018pib,Shanahan:2018nnv} for recent lattice studies.
In the former case, the proton has been the predominant object of study.
This is natural, as nucleons are often understood as the primary building
blocks of nuclear matter,
and understanding the mass and spin decomposition of the proton
is a necessary step in understanding the origin of most visible mass.
In the latter case, the pion is of great interest not only for the unique
role it plays in dynamical chiral symmetry breaking,
which is believed to be the origin of the majority of hadron mass,
but also because of its simplicity as a system and the ability to study
its properties (including components of its EMT) on the lattice~\cite{Horn:2016rip}.

On the other hand, the hadronic physics community has in general dedicated little attention to the internal structure of hadrons of spin higher than 1/2. From a theoretical point of view, a full picture of higher spin hadrons and nuclei is desirable because it would serve in elucidating QCD dynamics: spin-1 (and higher spin) systems carry information on non-nucleonic degrees of freedom, i.e.\ the dynamics beyond quarks and gluons confined within the individual nucleons~\cite{Jaffe:1989xy}.

Information on spin-1 hadrons  would allow us to thoroughly study such different parton contributions and dynamics in the spirit,  for instance, of the theoretical calculations of the gravitational
form factors for vector mesons in holographic QCD~\cite{Abidin:2008ku} and on the lattice~\cite{Detmold:2017oqb}. 
Being nearly the only experimentally available hadronic spin-1 target, the deuteron has  attracted a fair amount of attention over the past decades. It is the simplest bound state of more than one nucleon and, therefore, it has been of prime importance to unravel the nature of the nuclear binding. On the other hand, its internal structure and dynamics are the ultimate effect of the interactions between the elementary constituents, and this makes the deuteron a promising avenue towards understanding how QCD produces the force that binds nucleons together in nuclei~\cite{Boeglin:2015cha}.
After the first measurement by the HERMES collaboration of a tensor polarized collinear structure function of the deuteron~\cite{Airapetian:2005cb}, the so-called $b_1$ function defined in~\cite{Hoodbhoy:1988am}, it became clearer that going beyond the single-nucleon formulation is needed to describe the experimental data, especially in specific regions of the parton momentum range~\cite{Nikolaev:1996jy,Bora:1997pi,Edelmann:1997qe,Miller:2013hla,Cosyn:2017fbo}. 

The same arguments hold for the study of the gluonic content of higher spin hadrons, which requires once again to account for additional gluon functions in momentum and coordinate space that are exclusive to tensor polarized structures and therefore related to spin-1 or higher. This fact has stimulated a recent interest in the theoretical~\cite{Boer:2016xqr,Cotogno:2017puy} and lattice community~\cite{Detmold:2016gpy,Winter:2017bfs}. The deuteron is thus expected to play a major role in the 12 GeV program at Jefferson Lab (JLab) dedicated to spin-1 targets~\cite{Slifer:2014bda}.

As a fundamental entity encoding the spatial and mechanical
properties of hadrons,
the EMT of a spin-1 system such as the deuteron  
contains much of this dynamical information that is of interest
to the nuclear physics community.
This information is encoded in GFFs
familiar from the EMT of spin-$0$ and spin-$1/2$ systems,
but also within a host of additional form factors
novel to spin-$1$ systems.
This is analogous to the spin-$1$ electromagnetic current containing
one more form factor than the spin-$1/2$ current,
and has a similar origin.
A spin-$1$ system has an additional degree of freedom,
which can manifest itself in higher multipole moments
(in this case, a quadrupole moment) or a tensor polarization mode.
A full understanding of spin-$1$ systems requires a complete
categorization of all the independent Lorentz
structures that can appear in its EMT,
and an elucidation of the physical significance of the
GFFs that appear with these structures.

Expressions for the decomposition of the EMT for spin-$1$ hadrons have appeared earlier in Refs.~\cite{Holstein:2006ud,Abidin:2008ku,Taneja:2011sy}. In the present work, we provide the complete EMT decomposition that also includes all non-conserved terms (appearing incompletely in~\cite{Taneja:2011sy}) and we study the properties of and the relations between the GFFs that parametrize the local operator for the EMT. More specifically, we derive new sum rules and present expressions for the mass and angular momentum decomposition of a spin-$1$ hadron in terms of the new structures for quarks and gluons.

The results in this paper may be relevant for the experiments at JLab and a
future EIC~\cite{Accardi:2012qut} and for the proposed fixed-target projects $@$LHC~\cite{Kikola:2017hnp}, where different polarized hadrons and nuclei can be employed.  Current data for spin-$1$ tomography is rather scarce, with HERMES having measured deeply virtual Compton scattering (DVCS) on the deuteron with both unpolarized~\cite{Airapetian:2009bm} and polarized targets~\cite{Airapetian:2010aa}.  In these measurements, hadrons were not detected in the final state, but simulations were used to select 
a sample of enhanced coherent deuteron contribution.  More recently, Jefferson Lab has measured deeply virtual $\pi^0$ production on the deuteron~\cite{Mazouz:2017skh} and a recent letter of intent allows for coherent deuteron DVCS measurements~\cite{LOI_deutdvcs_coh}. The latter should also be possible in Hall B using the ALERT detector~\cite{Armstrong:2017wfw}. Finally, generalized distribution amplitudes (GDAs) for the rho-rho meson pair, accessible in the crossed reaction $\gamma^*\gamma\to\rho\rho$~\cite{Achard:2003qa,Anikin:2003fr,Anikin:2005ur}, can be related to the rho-meson GFFs similarly to the pion case~\cite{Kumano:2017lhr} and could also potentially be studied at Belle II. 

This work is organized as follows.
In Section~\ref{sec:decomp}, we give a full decomposition
of the most general form that the EMT of a spin-1 hadron can take.
This section also contains sum rules that follow immediately
from energy-momentum conservation.
In Section~\ref{sec:multipole}, we calculate the multipole moments
of matrix elements of the spin-1 EMT.
Mass and angular momentum decompositions are derived in this section,
along with additional sum rules and a work-energy balance relation.
Section~\ref{sec:gpd_links} explores the connections between the EMT
and Mellin moments of twist-2 and twist-3 generalized parton distributions.
Finally, in Section~\ref{sec:concl}, we summarize our results.
In addition, in Appendix~\ref{AppA} the form factors counting technique is reviewed, Appendix~\ref{AppB} and~\ref{AppC} contain additional information on Lorentz projectors and the polarization bilinears useful to obtain the parametrizations of the EMT, and Appendix~\ref{AppD} displays the covariant parametrization of the GPD correlator.

\section{Decomposition of the energy-momentum tensor}
\label{sec:decomp}
The goal of this section is to construct the most general possible
parametrization for the EMT of an on-shell, spin-1
hadron in terms of GFFs.
A variety of definitions exists for the EMT in QCD
(for a review, see \cite{Leader:2013jra}),
but here we work with the gauge-invariant kinetic form of the QCD EMT
$T^{\mu\nu}=T^{\mu\nu}_q+T^{\mu\nu}_g$, where
\begin{equation}
  T^{\mu\nu}_q=\frac{1}{2}\overline\psi\gamma^\mu
  i\overset{\leftrightarrow}{D}\!\!\!\!\phantom{D}^\nu\psi
  -g^{\mu\nu}\overline\psi\left(
  \frac{i}{2}\overset{\leftrightarrow}{\slashed{D}}\!\!\!\!\phantom{D}
  -m
  \right)\psi
  ,
  \qquad T^{\mu\nu}_g
  =
  -2\text{Tr}[F^{\mu\lambda}F^\nu_{\phantom{\nu}\lambda}]
  +\frac{1}{2}g^{\mu\nu}\text{Tr}[F^{\alpha\beta}F_{\alpha\beta}]
\end{equation}
with
$
  \overset{\leftrightarrow}{D}\!\!\!\!\phantom{D}^\mu
  =
  (\overset{\rightarrow}{\partial}\!\!\!\!\phantom{\partial}^\mu
  - \overset{\leftarrow}{\partial}\!\!\!\!\phantom{\partial}^\mu)
  -2igA^\mu
$.
Due to the presence of spin, the QCD EMT is in general not symmetric under exchange of its free Lorentz indices,
with the entirety of the asymmetry in the quark contribution.
The EMT $T^{\mu\nu}$ is a conserved current,
with the symmetric and antisymmetric components being separately
conserved.
Accordingly, we consider the general form of the EMT in two layers:
the symmetric component of the EMT and the full asymmetric EMT.

%%%%%%%%%%%%%%%%%%%%%%%%%%%%%%%%%%%%%%%%%%%%%%%%%%%%%%%%%%%%%%%%%%%%%%%%%%%%%%%%

\subsection{Symmetric EMT}

For a spin-1 system, there are only six possible independent rank-2 Lorentz structures that
are symmetric, $\mathsf P$-even, $\mathsf T$-even,
consistent with the hermiticity property, Lorentz-covariant,
linear in each of the initial and final state polarization vectors,
and conserved~\cite{Holstein:2006ud,Abidin:2008ku}.
This is fewer than the seven Lorentz structures that arise from the
$(1,1)$ representation of the Lorentz group
(see Appendix~\ref{AppA}),
meaning one of those Lorentz structures is non-conserved.
The conserved symmetric EMT takes the following form
\begin{align}
  \langle p^\prime, \lambda^\prime \mid T_{\mu\nu}(0) \mid p, \lambda \rangle
  &=
  % G1 ~~~~~~~~~~~~~~~~~~~~~~~~~~~~~~~~~~~~~~~~~~~~~~~~~~~~~~~~~~~~~~~~~~~~~~~~~
  -2P_\mu P_\nu\left[(\epsilon^{\prime*} \epsilon)
  \mathcal{G}_1(t)
  % G2 ~~~~~~~~~~~~~~~~~~~~~~~~~~~~~~~~~~~~~~~~~~~~~~~~~~~~~~~~~~~~~~~~~~~~~~~~~
  -
  \frac{(\Delta\epsilon^{\prime*} )(\Delta\epsilon )}{2M^2}
  \mathcal{G}_2(t)\right]
  % break
  \notag \\ &
  % G3 ~~~~~~~~~~~~~~~~~~~~~~~~~~~~~~~~~~~~~~~~~~~~~~~~~~~~~~~~~~~~~~~~~~~~~~~~~
  - \frac{1}{2}(\Delta_\mu \Delta_\nu - \Delta^2 g_{\mu\nu})
  \left[(\epsilon^{\prime*} \epsilon)
  \mathcal{G}_3(t)
  % G4 ~~~~~~~~~~~~~~~~~~~~~~~~~~~~~~~~~~~~~~~~~~~~~~~~~~~~~~~~~~~~~~~~~~~~~~~~~
  -
  \frac{(\Delta\epsilon^{\prime*} )(\Delta\epsilon )}{2M^2}
  \mathcal{G}_4(t)\right]
  % G5 ~~~~~~~~~~~~~~~~~~~~~~~~~~~~~~~~~~~~~~~~~~~~~~~~~~~~~~~~~~~~~~~~~~~~~~~~~
  +
  P_{\{\mu}\left( \epsilon^{\prime*}_{\nu\}} (\Delta \epsilon)
  - \epsilon_{\nu\}} (\Delta \epsilon^{\prime*}) \right)
  \mathcal{G}_5(t)
  % break
  \notag \\ &
  % G6 ~~~~~~~~~~~~~~~~~~~~~~~~~~~~~~~~~~~~~~~~~~~~~~~~~~~~~~~~~~~~~~~~~~~~~~~~~
  +
  \frac{1}{2} \left[
    \Delta_{\{\mu}\left( \epsilon^{\prime*}_{\nu\}} (\Delta\epsilon)
    + \epsilon_{\nu\}} (\Delta\epsilon^{\prime*}) \right) 
    - \epsilon^{\prime*}_{\{\mu}\epsilon_{\nu\}} \Delta^2
    - g_{\mu\nu}(\Delta\epsilon^{\prime*})(\Delta\epsilon)
    \right]
  \mathcal{G}_6(t)
  \label{eqn:EMT:symmetric:total}
  ,
\end{align}
where $M$ is the hadron mass, $P=(p'+p)/2$ is the average four-momentum, $t=\Delta^2$ with
$\Delta=p'-p$ is the four-momentum transfer,
and for each four-vector $a,b$ one has $a_{\{\mu} b_{\nu\}}=(a_\mu b_\nu+a_\nu b_\mu)/2$.
Energy and momentum must be conserved in a closed system,
so in this decomposition of the symmetric EMT a
sum over all partons is understood.
The partial EMT for quarks and gluons does not have to be conserved however,
so there are three additional independent Lorentz structures that can appear
\begin{align}
  \langle p^\prime, \lambda^\prime \mid T^a_{\mu\nu}(0) \mid p, \lambda \rangle
  &=
 % G1 ~~~~~~~~~~~~~~~~~~~~~~~~~~~~~~~~~~~~~~~~~~~~~~~~~~~~~~~~~~~~~~~~~~~~~~~~~
  -2P_\mu P_\nu\left[(\epsilon^{\prime*} \epsilon)
  \mathcal{G}^a_1(t)
  % G2 ~~~~~~~~~~~~~~~~~~~~~~~~~~~~~~~~~~~~~~~~~~~~~~~~~~~~~~~~~~~~~~~~~~~~~~~~~
  -
  \frac{(\Delta\epsilon^{\prime*} )(\Delta\epsilon )}{2M^2}
  \mathcal{G}^a_2(t)\right]
  % break
  \notag \\ &
  % G3 ~~~~~~~~~~~~~~~~~~~~~~~~~~~~~~~~~~~~~~~~~~~~~~~~~~~~~~~~~~~~~~~~~~~~~~~~~
  - \frac{1}{2}(\Delta_\mu \Delta_\nu - \Delta^2 g_{\mu\nu})
  \left[(\epsilon^{\prime*} \epsilon)
  \mathcal{G}^a_3(t)
  % G4 ~~~~~~~~~~~~~~~~~~~~~~~~~~~~~~~~~~~~~~~~~~~~~~~~~~~~~~~~~~~~~~~~~~~~~~~~~
  -
  \frac{(\Delta\epsilon^{\prime*} )(\Delta\epsilon )}{2M^2}
  \mathcal{G}^a_4(t)\right]
  % G5 ~~~~~~~~~~~~~~~~~~~~~~~~~~~~~~~~~~~~~~~~~~~~~~~~~~~~~~~~~~~~~~~~~~~~~~~~~
  +
  P_{\{\mu}\left( \epsilon^{\prime*}_{\nu\}} (\Delta \epsilon)
  - \epsilon_{\nu\}} (\Delta \epsilon^{\prime*}) \right)
  \mathcal{G}^a_5(t)
  % break
  \notag \\ &
  % G6 ~~~~~~~~~~~~~~~~~~~~~~~~~~~~~~~~~~~~~~~~~~~~~~~~~~~~~~~~~~~~~~~~~~~~~~~~~
  +
  \frac{1}{2} \left[
    \Delta_{\{\mu}\left( \epsilon^{\prime*}_{\nu\}} (\Delta\epsilon)
    + \epsilon_{\nu\}} (\Delta\epsilon^{\prime*}) \right) 
    - \epsilon_{\{\mu}^{\prime*}\epsilon_{\nu\}} \Delta^2
    - g_{\mu\nu}(\Delta\epsilon^{\prime*})(\Delta\epsilon)
    \right]
  \mathcal{G}^a_6(t)
  \notag \\ &
  % G7, G8, G9 ~~~~~~~~~~~~~~~~~~~~~~~~~~~~~~~~~~~~~~~~~~~~~~~~~~~~~~~~~~~~~~~~~
  +\epsilon_{\{\mu}^{\prime*}\epsilon_{\nu\}} M^2  \mathcal{G}^a_7(t)
  + g_{\mu\nu} M^2 (\epsilon'^*\epsilon) \mathcal{G}^a_8(t)
  + \frac{1}{2}g_{\mu\nu}(\Delta\epsilon'^*)(\Delta\epsilon) \mathcal{G}^a_9(t)
  ,
\end{align}
where $a=q,g$. Summing over all partons, we should recover~\eqref{eqn:EMT:symmetric:total} which implies the following sum rules
\begin{equation}\label{conservation}
      \sum_{a=q,g} \mathcal{G}_i^a(t) = 0\qquad\text{for}\quad i=7,8,9.
\end{equation}
Note that we have named the GFFs to agree with the conventions
in \cite{Taneja:2011sy},
although we find an additional non-conserved pure trace GFF,
in agreement with the counting in Appendix~\ref{AppA}. It should also be noted that unlike the total GFFs $\mathcal G_i(t)$, the partial GFFs $\mathcal G_i^a(t)$ are usually scale and scheme dependent.
%%%%%%%%%%%%%%%%%%%%%%%%%%%%%%%%%%%%%%%%%%%%%%%%%%%%%%%%%%%%%%%%%%%%%%%%%%%%%%%%

\subsection{Asymmetric EMT}

When the constituents of a system possess intrinsic angular momentum,
the EMT is in general expected to be asymmetric,
see e.g.~\cite{Leader:2013jra,Lorce:2018zpf,Florkowski:2018fap}
for recent discussions.
This is a simple consequence of the conservation of the
generalized angular momentum $\partial_\mu M^{\mu\alpha\beta}=0$,
with
$M^{\mu\alpha\beta}=x^\alpha T^{\mu\beta}-x^\beta T^{\mu\alpha}+S^{\mu\alpha\beta}$
and where $S^{\mu\alpha\beta}$ is the intrinsic generalized angular momentum tensor,
which when combined with the conservation of the EMT implies that
$T^{\alpha\beta}-T^{\beta\alpha}=-\partial_\mu S^{\mu\alpha\beta}$.
In agreement with the counting in Appendix~\ref{AppA},
we find only two antisymmetric Lorentz structures satisfying all the constraints.
The most general form of the EMT is therefore
\begin{align}
  \langle p^\prime, \lambda^\prime \mid T^a_{\mu\nu}(0) \mid p, \lambda \rangle
  &=
 % G1 ~~~~~~~~~~~~~~~~~~~~~~~~~~~~~~~~~~~~~~~~~~~~~~~~~~~~~~~~~~~~~~~~~~~~~~~~~
  -2P_\mu P_\nu\left[(\epsilon^{\prime*} \epsilon)
  \mathcal{G}^a_1(t)
  % G2 ~~~~~~~~~~~~~~~~~~~~~~~~~~~~~~~~~~~~~~~~~~~~~~~~~~~~~~~~~~~~~~~~~~~~~~~~~
  -
  \frac{(\Delta\epsilon^{\prime*} )(\Delta\epsilon )}{2M^2}
  \mathcal{G}^a_2(t)\right]
  % break
  \notag \\ &
  % G3 ~~~~~~~~~~~~~~~~~~~~~~~~~~~~~~~~~~~~~~~~~~~~~~~~~~~~~~~~~~~~~~~~~~~~~~~~~
  - \frac{1}{2}(\Delta_\mu \Delta_\nu - \Delta^2 g_{\mu\nu})
  \left[(\epsilon^{\prime*} \epsilon)
  \mathcal{G}^a_3(t)
  % G4 ~~~~~~~~~~~~~~~~~~~~~~~~~~~~~~~~~~~~~~~~~~~~~~~~~~~~~~~~~~~~~~~~~~~~~~~~~
  -
  \frac{(\Delta\epsilon^{\prime*} )(\Delta\epsilon )}{2M^2}
  \mathcal{G}^a_4(t)\right]
  % G5 ~~~~~~~~~~~~~~~~~~~~~~~~~~~~~~~~~~~~~~~~~~~~~~~~~~~~~~~~~~~~~~~~~~~~~~~~~
  +
  P_{\{\mu}\left( \epsilon^{\prime*}_{\nu\}} (\Delta \epsilon)
  - \epsilon_{\nu\}} (\Delta \epsilon^{\prime*}) \right)
  \mathcal{G}^a_5(t)
  % break
  \notag \\ &
  % G6 ~~~~~~~~~~~~~~~~~~~~~~~~~~~~~~~~~~~~~~~~~~~~~~~~~~~~~~~~~~~~~~~~~~~~~~~~~
  +
  \frac{1}{2} \left[
    \Delta_{\{\mu}\left( \epsilon^{\prime*}_{\nu\}} (\Delta\epsilon)
    + \epsilon_{\nu\}} (\Delta\epsilon^{\prime*}) \right) 
    - \epsilon_{\{\mu}^{\prime*}\epsilon_{\nu\}} \Delta^2
    - g_{\mu\nu}(\Delta\epsilon^{\prime*})(\Delta\epsilon)
    \right]
  \mathcal{G}^a_6(t)
  \notag \\ &
  % G7, G8, G9 ~~~~~~~~~~~~~~~~~~~~~~~~~~~~~~~~~~~~~~~~~~~~~~~~~~~~~~~~~~~~~~~~~
  +\epsilon_{\{\mu}^{\prime*}\epsilon_{\nu\}} M^2  \mathcal{G}^a_7(t)
  + g_{\mu\nu} M^2 (\epsilon'^*\epsilon) \mathcal{G}^a_8(t)
  + \frac{1}{2}g_{\mu\nu}(\Delta\epsilon'^*)( \Delta\epsilon)   \mathcal{G}^a_9(t)
   \notag \\ &
  % G10, G11 ~~~~~~~~~~~~~~~~~~~~~~~~~~~~~~~~~~~~~~~~~~~~~~~~~~~~~~~~~~~~~~~~~~~
  +
  P_{[\mu}\left( \epsilon^{\prime*}_{\nu]} (\Delta \epsilon)
  - \epsilon_{\nu]} (\Delta \epsilon^{\prime*}) \right)
  \mathcal{G}^a_{10}(t)
  +
  \Delta_{[\mu}\left( \epsilon^{\prime*}_{\nu]} (\Delta \epsilon)
  + \epsilon_{\nu]} (\Delta \epsilon^{\prime*}) \right)
  \mathcal{G}^a_{11}(t),
  \label{eqn:EMT:general}
\end{align}
where $a_{[\mu} b_{\nu]}=(a_\mu b_\nu-a_\nu b_\mu)/2$.
Since one of the two new tensors is non-conserved,
energy-momentum conservation imposes the additional sum rule
\begin{equation}
    \sum_{a=q,g} \mathcal{G}_{11}^a(t) = 0.
\end{equation}
This is an interesting new feature of the spin-1 target,
since a spin-0 target has no antisymmetric part and a spin-1/2 target has
only a conserved contribution.
This has to do with the fact that
the intrinsic generalized angular momentum tensor for a scalar field vanishes,
$S^{\mu\alpha\beta}_0=0$,
and is completely antisymmetric for a Dirac field, as
$
  S^{\mu\alpha\beta}_{1/2}
  =
  \frac{1}{2}\,\epsilon^{\mu\alpha\beta\lambda}
  \overline\psi\gamma_\lambda\gamma_5\psi
$ with $\epsilon_{\,0123}=+1$.
In the case of a massive vector field $V^\mu$,
the intrinsic generalized angular momentum tensor reads $S^{\mu\alpha\beta}_1=-2F^{\mu[\alpha}V^{\beta]}$,
so that $\partial_\alpha\partial_\mu S^{\mu\alpha\beta}_1\neq 0$,
opening the possibility of having a non-vanishing intrinsic energy
dipole moment beside intrinsic angular momentum~\cite{Lorce:2018zpf}. 

Since we will be working with the kinetic form~\cite{Leader:2013jra} of the QCD EMT,
the QCD equations of motion imply that the antisymmetric part of
the EMT can be expressed in terms of the axial-vector current
as follows~\cite{Balitsky:1987bk,Leader:2013jra,Lorce:2015lna}
\begin{equation}\label{QCDid}
\overline\psi\gamma^{[\mu}i\overset{\leftrightarrow}{D}\!\!\!\!\phantom{D}^{\nu]}\psi=-\frac{1}{2}\epsilon^{\mu\nu\rho\sigma}\partial_\rho(\overline\psi\gamma_\sigma\gamma_5\psi).
\end{equation}
The matrix elements of the axial-vector current being parametrized as~\cite{Frederico:1992vm,Berger:2001zb}
\begin{equation}
\langle p^\prime, \lambda^\prime \mid \overline\psi(0)\gamma_\mu\gamma_5\psi(0) \mid p, \lambda \rangle=-2i\epsilon_{\mu\alpha\beta P}\left(\epsilon^{\prime*\alpha}\epsilon^\beta\tilde G_1(t)+\frac{\Delta^\alpha[\epsilon^{\prime*\beta}(\Delta\epsilon)-\epsilon^\beta(\Delta\epsilon^{\prime *})]}{M^2}\tilde G_2(t)\right)
\end{equation}
with the notation $\epsilon_{\mu\alpha\beta P}=\epsilon_{\mu\alpha\beta \lambda}P^\lambda$, we find from considering the matrix elements of~\eqref{QCDid}
\begin{subequations}
\begin{align}
\mathcal{G}_{10}^q(t) &= -\tilde G_1(t)+\frac{t}{M^2}\tilde G_2(t),\\
\mathcal{G}_{11}^q(t) &= 0,\label{G11}
\end{align}  
\end{subequations}
and $\mathcal{G}_{10}^g(t)=\mathcal{G}_{11}^g(t)=0$. Note that the vanishing of the antisymmetric part of $T^g_{\mu\nu}$ has to do with the impossibility of writing down the gluon spin contribution in a form that is both local and gauge invariant~\cite{Leader:2013jra}.
We thus find that the antisymmetric part of the EMT for a spin-1 hadron
is conserved\footnote{
  We note that this conclusion holds for the gauge-invariant kinetic EMT,
  but may not hold for the (non-gauge-invariant) canonical EMT.
  This is because the gluon contribution to the canonical EMT is asymmetric,
  and due to the fact that
  ${\partial_\alpha\partial_\mu S^{\mu\alpha\beta}_1\neq 0}$,
  the divergences of the symmetric and antisymmetric parts
  of the canonical gluon EMT are not expected to separately vanish.
}.

\section{Multipole moments of the energy-momentum tensor}
\label{sec:multipole}

Much of the interesting information about a hadron's mechanical properties
that is contained in the EMT is encoded by the multipole moments of
the EMT matrix elements.
These include static observables such as mass, angular momentum,
the inertia tensor, and so on, but additionally include information
about how each of these decomposes into quark and gluon contributions.
The decomposition of hadron mass and angular momentum
into quark and gluon contributions---and
the latter also into spin and orbital angular momentum components---has
been a major focus of recent literature on the EMT.
As with much of the other literature on the QCD EMT,
this focus has been primarily directed towards spin-1/2 systems
(predominantly the proton) and spin-0.
In this section, we elaborate on the mechanical properties of spin-1
hadrons encoded by the multipole moments of their EMT,
including both properties that are analogous to the lesser-spin cases
and those that are new to spin-1.

\subsection{Mass decomposition and balance equation}

The mass decomposition and balance equation associated with a spin-1 target are obtained in terms of the following properly normalized matrix element of the EMT~\cite{Lorce:2017xzd,Lorce:2018egm}
\begin{equation}
\langle\langle\int\textrm{d}^3r\, T^a_{\mu\nu}(0,\boldsymbol r)\rangle\rangle\equiv \frac{\langle p, \lambda^\prime \mid \int\textrm{d}^3r\, T^a_{\mu\nu}(0,\vec r) \mid p, \lambda \rangle}{\langle p, \lambda\mid p, \lambda \rangle}=\frac{1}{2p^0}\,\langle p, \lambda^\prime \mid T^a_{\mu\nu}(0) \mid p, \lambda \rangle.
\end{equation}
Using the covariant expression for the density matrix of a spin-1 system~\cite{Joos:1964,Zwanziger:1965} (see also App.~\ref{AppC} for more details)
\begin{equation}
\epsilon_\beta\epsilon^*_\alpha=-\frac{1}{3}P_{\beta\alpha}+\frac{i}{2M}\epsilon_{\beta\alpha\mathcal Sp}-\mathcal T_{\beta\alpha},  
\end{equation}
where the projector onto the subspace orthogonal to $p^\mu$ is given by
\begin{equation}
P_{\mu\nu}=g_{\mu\nu}-\frac{p_\mu p_\nu}{M^2},
\end{equation}
and the covariant vector and tensor polarizations by
\begin{align}
\mathcal S^\mu(p)&=-\epsilon^{\mu\rho\sigma\lambda}\text{Im}(\epsilon_\rho\epsilon^*_\sigma)\frac{p_\lambda}{M},\\
\mathcal T_{\mu\nu}(p)&=-\frac{1}{3}P_{\mu\nu}-\text{Re}(\epsilon_\mu\epsilon^*_\nu),
\end{align}
we find
\begin{equation}\label{FL}
\langle p, \lambda^\prime \mid T^a_{\mu\nu}(0) \mid p, \lambda \rangle
= 2p_\mu p_\nu\left[\mathcal G^a_1(0)+\frac{1}{6}\mathcal G^a_7(0)\right]-2g_{\mu\nu}M^2\left[\frac{1}{2}\mathcal G^a_8(0)+\frac{1}{6}\mathcal G^a_7(0)\right]-\mathcal T_{\mu\nu} M^2\mathcal G^a_7(0).
\end{equation}
The first two Lorentz structures do not depend on the spin and are indeed common to all targets. The last Lorentz structure is new. It is related to the target tensor polarization and therefore does not appear in the case of spin-0 or spin-1/2 targets. Because of Poincar\'e invariance, the forward matrix element of the total EMT has to assume the form\footnote{This is consistent with the total four-mometum of the system being given by $p^\mu=\langle\langle\int\textrm{d}^3r\, T^{0\mu}(0,\vec r)\rangle\rangle$.}
\begin{equation}
\langle p, \lambda^\prime \mid T_{\mu\nu}(0) \mid p, \lambda \rangle
= 2p_\mu p_\nu,
\end{equation}
from which we conclude that
\begin{equation}
    \sum_{a=q,g} \mathcal{G}_1^a(0) = 1
\end{equation}
using the constraints in Eq.~\eqref{conservation}.

The Lorentz-invariant coefficients in Eq.~\eqref{FL} can be interpreted in terms of proper internal energy and pressure-volume work~\cite{Lorce:2017xzd,Lorce:2018egm}. In the target rest frame, the partial internal energy is given by\footnote{One can of course avoid having recourse to  the rest frame and use instead the projectors $p^\alpha p^\beta/M^2$ and $P_{\alpha\beta}$~\cite{Lorce:2017xzd}.}
\begin{equation}
U_a=\frac{1}{2M}\langle p, \lambda^\prime \mid T_a^{00}(0) \mid p, \lambda \rangle
=\left[\mathcal G^a_1(0)-\frac{1}{2}\mathcal G^a_8(0)\right]M,
\end{equation}
and the partial isotropic pressure-volume work by
\begin{equation}
W_a=\frac{\delta^{ij}}{6M}\langle p, \lambda^\prime \mid T_a^{ij}(0) \mid p, \lambda \rangle
=\left[\frac{1}{2}\mathcal G^a_8(0)+\frac{1}{6}\mathcal G^a_7(0)\right]M.
\end{equation}
The new feature of a spin-1 target is the presence of a partial pressure-volume work anisotropy
\begin{equation}
W^{ij}_a=\frac{1}{2M}\langle p, \lambda^\prime \mid T_a^{ij}(0) \mid p, \lambda \rangle
-\delta^{ij}W_a=\mathcal T^{ij}\left[-\frac{1}{2}\mathcal G^a_7(0)\right]M
\end{equation}
associated with the tensor polarization. The mass decomposition then takes the form
\begin{equation}
M=\sum_{a=q,g}U_a
\end{equation}
and the balance equations read
\begin{equation}
\sum_{a=q,g} W_a=0,\qquad\sum_{a=q,g} W^{ij}_a=0.
\end{equation}

\subsection{Angular momentum decomposition}

As explained in detail in~\cite{Bakker:2004ib,Leader:2013jra}, higher spatial moments of the energy-momentum distribution are ambiguous if defined naively as $\langle\langle\int\textrm{d}^3r\,r^j T^a_{\mu\nu}(0,\vec r)\rangle\rangle$. The reason for this is because information about the spatial distribution is lost in the forward limit $\Delta\to 0$. Spatial distributions can only be defined in frames where no energy is transfered to the system $\Delta^0=\vec P\cdot\vec\Delta/P^0=0$. In this work, we will only consider the Breit frame $\vec P=\vec 0$ where three-dimensional spatial distributions of the EMT are defined as~\cite{Polyakov:2002yz,Lorce:2017wkb,Lorce:2018egm}
\begin{equation}
\langle T_a^{\mu\nu}\rangle(\vec r)\equiv\int\frac{\textrm{d}^3\Delta}{(2\pi)^3}\,e^{-i\vec\Delta\cdot\vec r}\,\frac{1}{2P^0}\langle \frac{\vec\Delta}{2}, \lambda^\prime \mid T_a^{\mu\nu}(0) \mid -\frac{\vec\Delta}{2}, \lambda \rangle
\end{equation}
with $P^0=\sqrt{M^2+\frac{\vec\Delta^2}{4}}$. The dipole moment of the spatial distribution is then given by
\begin{equation}
\int\textrm{d}^3r\,r^j \langle T_a^{\mu\nu}\rangle(\vec r)=\left\{-i\nabla^j_\Delta\left[\frac{1}{2P^0}\langle \frac{\vec\Delta}{2}, \lambda^\prime \mid T_a^{\mu\nu}(0) \mid -\frac{\vec\Delta}{2}, \lambda \rangle\right]\right\}_{\Delta=0}.
\end{equation}
Using the Breit-frame expansion of the polarization four-vector bilinear derived in Appendix~\ref{AppC}, we find
\begin{equation}\label{DipoleBF}
\int\textrm{d}^3r\,r^j \langle T_a^{\mu\nu}\rangle(\vec r)=\frac{1}{2}g^{0\{\mu}\epsilon^{\nu\} j\mathcal S 0}\left[\mathcal G^a_5(0)+\frac{1}{2}\mathcal G^a_7(0)\right]+\frac{1}{2}g^{0[\mu}\epsilon^{\nu] j\mathcal S 0}\,\mathcal G^a_{10}(0).
\end{equation}
Clearly, the only non-vanishing dipole moment is associated with the momentum distribution $\langle T_a^{0k}\rangle(\vec r)$ and is orthogonal to the vector polarization of the target $\mathcal S^\mu=(0,\vec s)$. It simply originates from the parton orbital angular momentum (OAM)\footnote{Since we consider the local gauge-invariant EMT, we are dealing with the kinetic form of OAM, see e.g.~\cite{Leader:2013jra,Liu:2015xha} for more details.}
\begin{equation}\label{Lq}
L^i_a=\epsilon^{ijk}\int\textrm{d}^3r\,r^j \langle T_a^{0k}\rangle(\vec r)=\frac{s^i}{2}\left[\mathcal G^a_5(0)+\frac{1}{2}\mathcal G^a_7(0)+\mathcal G^a_{10}(0)\right].
\end{equation}
In QCD, we then find that the parton total angular momentum (AM) is given in the target rest frame by
\begin{equation}\label{Jq}
J^i_a=\epsilon^{ijk}\int\textrm{d}^3r\,r^j \langle T_a^{\{0k\}}\rangle(\vec r)=\frac{s^i}{2}\left[\mathcal G^a_5(0)+\frac{1}{2}\mathcal G^a_7(0)\right].
\end{equation}
We naturally recover $J^i_q=L^i_q+S^i_q$ with the quark spin contribution being given by
\begin{equation}
    S^i_q=\frac{1}{2}\int\textrm{d}^3r\,\langle \overline\psi\gamma^i\gamma_5\psi\rangle(\vec r)=\frac{s^i}{2}\tilde G_1(0)=-\frac{s^i}{2}\mathcal G^q_{10}(0).
\end{equation}
For gluons we simply have $J^i_g=L^i_g$ because no local gauge-invariant definition of the gluon spin does exist, see e.g.~\cite{Leader:2013jra} for a recent detailed discussion. 

Beside the term proportional to $\mathcal G^a_5$ already obtained in~\cite{Taneja:2011sy}, we also find a contribution from the $\mathcal G^a_7$ GFF describing the non-conserved part of the EMT. Interestingly, such a contribution cannot appear for spin-0 and 1/2 targets, since in these cases the non-conserved terms are necessarily of the form of a pure trace and hence decoupled from AM. Summing over all the partons, the $\mathcal G_7$ contribution drops out according to Eq.~\eqref{conservation} and we get the AM constraint
\begin{equation}
    \sum_{a=q,g} \mathcal{G}_5^a(0) = 2
\end{equation}
 first derived in~\cite{Abidin:2008ku}.
 
\subsection{Mass radius and inertia tensor}

Beside mass, two other important quantities characterizing the energy distribution can be defined, namely the mass radius and the inertia tensor. Both are expressed in terms of the following second-order moments
\begin{equation}
 C^{ij}_a=\int\textrm{d}^3r\,r^i r^j \langle T_a^{00}\rangle(\vec r)=\left\{-\nabla^i_\Delta\nabla^j_\Delta\left[\frac{1}{2P^0}\langle \frac{\vec\Delta}{2}, \lambda^\prime \mid T_a^{00}(0) \mid -\frac{\vec\Delta}{2}, \lambda \rangle\right]\right\}_{\Delta=0}.
\end{equation}
The mass radius defined as $R_M=\sqrt{\langle r^2\rangle}$ with
\begin{equation}
\langle r^2\rangle=\frac{1}{M}\sum_{a=q,g} C^{ij}_a\delta^{ij}
\end{equation}
gives an idea of the spatial extension of the energy distribution. The inertia tensor~\cite{keith,goldstein} defined as
\begin{equation}
 I^{ij}_a= C^{kl}_a\left(\delta^{kl}\delta^{ij}-\delta^{ik}\delta^{jl}\right)
\end{equation}
allows one to determine the moment of inertia $I_a^{\vec n}= I^{ij}_a n^in^j$ of the system about an arbitrary axis $\vec n$ passing through the center of mass, which coincides in the Breit frame with the origin~\cite{Lorce:2018zpf}. It is related to the mass quadrupole moment
\begin{equation}
Q^{ij}_a= C^{kl}_a\left(\delta^{ik}\delta^{jl}-\frac{1}{3}\delta^{kl}\delta^{ij}\right)=- I^{kl}_a\left(\delta^{ik}\delta^{jl}-\frac{1}{3}\delta^{kl}\delta^{ij}\right)
\end{equation}
which measures the deviation from a spherical distribution of the energy.

Using once more the Breit-frame expansion of the polarization four-vector bilinear derived in Appendix~\ref{AppC}, we find
\begin{equation}
 C^{ij}_a=\frac{1}{M}\left[\delta^{ij}\left(\mathcal A_a(0)+\frac{1}{3}\mathcal B_a(0)\right)-\mathcal T^{ij}\mathcal B_a(0)\right]
\end{equation}
with
\begin{subequations}
\begin{align}
\mathcal A_a(t)&=-\frac{1}{4}\left(\mathcal G^a_1(t)+2\mathcal G^a_3(t)+\frac{1}{2}\mathcal G^a_8(t)\right)+2M^2\frac{\mathrm d}{\mathrm dt}\left(\mathcal G^a_1(t)-\frac{1}{2}\mathcal G^a_8(t)\right),\\
\mathcal B_a(t)&=-\mathcal G^a_1(t)-\mathcal G^a_2(t)+\mathcal G^a_5(t)+\frac{1}{2}\mathcal G^a_6(t)+\frac{1}{4}\mathcal G^a_7(t)+\frac{1}{2}\mathcal G^a_8(t)-\frac{1}{2}\mathcal G^a_9(t).  
\end{align}
\end{subequations}
The squared mass radius, inertia tensor and mass quadrupole moment are then given by
\begin{align}
\langle r^2\rangle&=\frac{1}{M^2}\sum_{a=q,g}\left[3\mathcal A_a(0)+\mathcal B_a(0)\right],\\
 I^{ij}_a&=\frac{1}{M}\left[2\delta^{ij}\left(\mathcal A_a(0)+\frac{1}{3}\mathcal B_a(0)\right)+\mathcal T^{ij}\mathcal B_a(0)\right],\\
Q^{ij}_a&=-\frac{1}{M}\mathcal T^{ij}\mathcal B_a(0).
\end{align}
As expected, the quadrupole moment in spin-1 hadrons is different from zero  due to the presence of the tensor polarization.

\section{GFFs as moments of GPDs}
\label{sec:gpd_links}
The connection between the EMT and
partonic distributions has long been a topic of consideration
(see e.g.~\cite{Jaffe:1989jz,Ji:1996ek,Ji:1998pc}).
Generalized parton distributions (GPDs) in particular allow
for GFFs to be extracted from their Mellin moments.
Since GPDs parametrize the non-perturbative structure contributing
to hard reactions such as deeply virtual compton scattering (DVCS)
and virtual meson production,
they are the most promising avenue for experimentally exploring
the form factors appearing in the EMT decomposition.

Leading-twist GPDs have been extensively studied for their
polynomiality relations~\cite{Ji:1998pc}.
A specific case allows the second Mellin moments of
helicity-independent twist-2 GPDs to be related to the
non-trace GFFs appearing in the symmetric component of the EMT.
Such relations have been studied not only for
spin-0~\cite{Polyakov:1999gs,Son:2014sna,Kumano:2017lhr}
and spin-1/2~\cite{Ji:1996ek,Ji:1998pc,Lorce:2018egm},
but also spin-1
systems~\cite{Abidin:2008ku,Taneja:2011sy,Detmold:2017oqb,Cosyn:2018thq}.

Beyond leading twist, the Penttinen-Polyakov-Shuvaev-Strikman (PPSS)
sum rule~\cite{Penttinen:2000dg}
relates the second moment of a twist-3 GPD
to the orbital angular momentum carried by quarks in spin-1/2 hadrons,
and it has also been shown \cite{Leader:2013jra}
that twist-4 GPDs contain information about the non-conserved
GFF in a nucleon.
In this section we will derive a spin-1 analogue of
the PPSS sum rule.

We proceed to derive sum rules for second Mellin moments of
GPDs up to twist-3\footnote{
  It is possible to consider up to twist-4 for quarks and twist-6 for gluons,
  but contributions beyond twist-3 are less promising for future
  phenomenological studies.
}.
Consider the quark and gluon vector correlators \footnote{The Ji convention~\cite{Ji:1998pc}
  is used in this work for the gluon GPDs
  to simplify all the following formulas.
  Note that
  ${2x H_{g,\mathrm{Ji}} = H_{g,\mathrm{Diehl}}}$~\cite{Diehl:2003ny}.
}
\begin{subequations}
\label{GPDcorrelators}
\begin{align}
  V_{\lambda\lambda^\prime}^{\mu,q}
  &=
  \langle p^\prime, \lambda^\prime |
  \mathcal{O}^\mu_{qV}
  | p, \lambda \rangle
  =
  \frac{1}{2}
  \int_{-\infty}^\infty \frac{\mathrm{d}\kappa}{2\pi}
  e^{i(Pn)\kappa x}
  \left\langle p^\prime, \lambda^\prime \middle|
  \overline{\psi}\!\left(-\frac{n\kappa}{2}\right)
  \gamma^\mu
  \left[-\frac{n\kappa}{2},\frac{n\kappa}{2}\right]
  \psi\!\left(\frac{n\kappa}{2}\right)
  \middle| p, \lambda \right\rangle,\label{GPDcorrelatorq}
  \\
  % ~~~~~~~~~~~~~~~~~~~~~~~~~~~~~~~~~~~~~~~~~~~~~~~~~~~~~~~~~~~~~~~~~~~~~~~~~~~~
  V_{\lambda\lambda^\prime}^{\mu,g}
  &=
  \langle p^\prime, \lambda^\prime |
  \mathcal{O}^\mu_{gV}
  | p, \lambda \rangle
  =
  \frac{\delta^{\phantom{\{}\mu}_{\{\alpha} n^{\phantom{\mu}}_{\beta\}}}{x(Pn)}
  \int_{-\infty}^\infty \frac{\mathrm{d}\kappa}{2\pi}
  e^{i(Pn)\kappa x}
  \left\langle p^\prime, \lambda^\prime \middle|
  \mathrm{Tr}\left\{
    \left[\frac{n\kappa}{2},-\frac{n\kappa}{2}\right]
    F^{\alpha\lambda}\!\left(-\frac{n\kappa}{2}\right)
    \left[-\frac{n\kappa}{2},\frac{n\kappa}{2}\right]
    F_\lambda^{\phantom{\lambda}\beta}\!\left(\frac{n\kappa}{2}\right)
    \right\}
  \middle| p, \lambda \right\rangle,
\end{align}
\end{subequations}
where $n$ is a light-like four-vector and $[y,z]$
denotes a straight Wilson line joining the spacetime points $y$ and $z$.
These correlators enter the description of deeply virtual Compton scattering
and can be parametrized up to twist 3 as
follows\footnote{For the complete covariant parametrization, see Appendix~\ref{AppD}.}~\cite{Berger:2001zb,Anikin:2011aa}
\begin{equation}
\begin{aligned}
    V_{\lambda \lambda'}^{\mu,a} &= 
    %%Twist 2
    %H1
    -\frac{P^\mu}{(Pn)} (\epsilon^{\prime*}\epsilon) \; H_1^a 
    %H2
    + \frac{\epsilon^{\prime*\mu} (\Delta \epsilon )-\epsilon^\mu (\Delta \epsilon^{\prime*} )}{2(Pn)}\, H_2^a
    %H3
    + \frac{P^\mu}{(Pn)}\frac{(\Delta \epsilon^{\prime*} )(\Delta \epsilon )}{2M^2} \, H_3^a
    %H4
    - \frac{\epsilon^{\prime*\mu} (\Delta \epsilon )+\epsilon^\mu (\Delta \epsilon^{\prime*} )}{2(Pn)}\,  H_4^a
     \\
    %--------
    %H5
    &\quad +\left[M^2 \frac{\epsilon^{\prime*\mu}(n \epsilon )+\epsilon^\mu (n \epsilon^{\prime*}) }{2(Pn)^2} + \frac{1}{3} \frac{P^\mu}{(Pn)}(\epsilon^{\prime*}\epsilon) \right]H_5^a
    %%Twist 3
    %G1 
    -\frac{\Delta^\mu_T}{(Pn)}(\epsilon^{\prime*}\epsilon) \; G_1^a 
    %G2
    -\frac{\Delta^\mu_T}{(Pn)}\frac{(n\epsilon^{\prime*}) (\Delta \epsilon )-(n\epsilon) (\Delta \epsilon^{\prime*} )}{2(Pn)}\, G_2^a
    \\
    %----------------
    %G3
    &\quad +\frac{\Delta^\mu_T}{(Pn)}\frac{(\Delta \epsilon^{\prime*} )(\Delta \epsilon )}{2M^2} \; G_3^a
    %G4
    -\frac{\Delta^\mu_T}{(Pn)}\frac{(n\epsilon^{\prime*}) (\Delta \epsilon )+(n\epsilon) (\Delta \epsilon^{\prime*} )}{2(Pn)}\, G_4^a
    %G5
    +\frac{\Delta^\mu_T}{(Pn)}\left[M^2 \frac{(n \epsilon^{\prime*})(n \epsilon )}{(Pn)^2} + \frac{1}{3}(\epsilon^{\prime*}\epsilon) \right]G_5^a
    \\
    %----------------
    %G6
    &\quad +\frac{\epsilon^{\prime*\mu}_T (\Delta \epsilon )-\epsilon^\mu_T (\Delta \epsilon^{\prime*} )}{2(Pn)} \, G_6^a
    %G7
    - \frac{\epsilon^{\prime*\mu}_T (\Delta \epsilon )+\epsilon^\mu_T (\Delta \epsilon^{\prime*} )}{2(Pn)}\, G_7^a
    %G8
    + M^2 \frac{\epsilon^{\prime*\mu}_T(n \epsilon )+\epsilon^\mu_T (n \epsilon^{\prime*}) }{2(Pn)^2}\, G_8^a
    %G9
    + M^2 \frac{\epsilon^{\prime*\mu}_T(n \epsilon )-\epsilon^\mu_T (n \epsilon^{\prime*}) }{2(Pn)^2}\, G_9^a .
    \label{eq:vectorcorr:twist3}
\end{aligned}
\end{equation}
The first five terms ($H^a_i$) correspond to the twist-2 GPDs,
and the remaining nine ($G^a_i$) are purely twist-3. In the quark sector, the twist-3 GPDs satisfy the relation
$\int\mathrm{d}x\,G_i^q=0$
as a consequence of the charge current conservation.
We suppressed the dependence of the GPDs on the parton longitudinal momentum $x$, longitudinal momentum transfer $\xi=-(\Delta n)/2(Pn)$, and squared momentum transfer $t=\Delta^2$ for conciseness of notation,
and made use of the Sudakov decomposition of four-vectors
$(n^2=\bar{n}^2=0,\, n\bar{n}=1)$
\begin{equation}
    z^\mu = (z n) \bar{n}^\mu + (z \bar{n}) n^\mu +  z_T^\mu.
\end{equation}

The second Mellin moment of the light-front string operators are related to the EMT up to twist 3 as follows
\begin{subequations}\label{eqn:gpd:moments}
\begin{align}
  \int_{-1}^1\mathrm{d}x\,
  x \mathcal{O}^\mu_{qV}
  &=
  \frac{
    1
  }{4(P  n)^{2}}
  \overline{\psi}(0)
  \gamma^\mu
  (i\overset{\leftrightarrow}{D}\!\!\!\!\phantom{D}n)
  \psi(0)
  =
  \frac{T_q^{\mu n}}{2(Pn)^2}, 
  \\
  \int_{-1}^1\mathrm{d}x\,x
  \mathcal{O}^\mu_{gV}
  &=
  \frac{
    1
  }{(Pn)^2}
  \mathrm{Tr}\left[
    F^{\mu\lambda}(0)
    F_{\lambda n}(0)
    \right]
  =
  \frac{T_g^{\mu n}}{2(Pn)^2} 
  .
\end{align}
\end{subequations}
Taking the off-forward matrix element on both sides allows us to relate seven of the GFFs to moments of leading-twist vector GPDs~\cite{Taneja:2011sy,Cosyn:2018thq}.
Comparing the Mellin moment of Eq.~\eqref{eq:vectorcorr:twist3} with the decomposition in
Eq.~\eqref{eqn:EMT:symmetric:total} in the symmetric frame $P^\mu_T=0$,
we find the following relations for quarks and gluons at twist 2
\begin{subequations}
\begin{align}
  % H1
  \int_{-1}^1\mathrm{d}x\, x \left[ H^{a}_1(x,\xi,t) - \frac{1}{3}H^{a}_5(x,\xi,t)\right] 
  &= \mathcal{G}^{a}_1(t) + \xi^2 \mathcal{G}^{a}_3(t),
  \\
  % H2
  \int_{-1}^1\mathrm{d}x\, x H^{a}_2(x,\xi,t)  &= \mathcal{G}^{a}_5(t),
  \\
  % H3
  \int_{-1}^1\mathrm{d}x\, x H^{a}_3(x,\xi,t)  &= \mathcal{G}^{a}_2(t)
  + \xi^2 \mathcal{G}^{a}_4(t),
  \\
  % H4
  \int_{-1}^1\mathrm{d}x\, x H^{a}_4(x,\xi,t)  &= \xi \mathcal{G}^{a}_6(t),
  \\
  % H5
  \int_{-1}^1\mathrm{d}x\, x H^{a}_5(x,\xi,t)  &= -\frac{t}{4M^2}\mathcal{G}^{a}_6(t)
  + \frac{1}{2}\mathcal{G}^{a}_7(t).
  \label{eqn:GPD_GFF:H5}
\end{align}
\end{subequations}
Since the GFFs appear as second Mellin moments of GPDs,
they are special cases of generalized form factors,
which correspond to arbitrary moments of GPDs.
We note the following correspondence between the GFFs
as defined in this work,
and the $s=2$ generalized form factors appearing in Ref.~\cite{Cosyn:2018thq}:
\begin{subequations}
\begin{align}
  A^{a}_{2,0}(t) &= \mathcal{G}^{a}_1(t), \\
  B^{a}_{2,0}(t) &= \mathcal{G}^{a}_5(t), \\
  C^{a}_{2,0}(t) &= \mathcal{G}^{a}_2(t), \\
  2 D^{a}_{2,1}(t) &= -\mathcal{G}^{a}_6(t), \\
  E^{a}_{2,1}(t) &= -\frac{t}{4M^2}\mathcal{G}^{a}_6(t) + \frac{1}{2}\mathcal{G}^{a}_7(t), \\
  4 F^{a}_{2}(t)   &= \mathcal{G}^{a}_3(t), \\
  4 H^{a}_{2}(t)   &= \mathcal{G}^{a}_4(t).
\end{align}
\end{subequations}
We also find the following relations for quarks at twist 3
\begin{subequations}
\begin{align}
  % G1
  \int_{-1}^1\mathrm{d}x\, x  G^{q}_1(x,\xi,t) 
  &= -\frac{\xi}{2} \mathcal{G}^{q}_3(t)=-\frac{1}{4}\frac{\partial}{\partial\xi}\int_{-1}^1\mathrm{d}x\, x \left[ H^{q}_1(x,\xi,t) - \frac{1}{3}H^{q}_5(x,\xi,t)\right],
  \\
  % G2
  \int_{-1}^1\mathrm{d}x\, x G^{q}_2(x,\xi,t)  &= 0,
  \\
  % G3
  \int_{-1}^1\mathrm{d}x\, x G^{q}_3(x,\xi,t)  &= -\frac{\xi}{2} \mathcal{G}^{q}_4(t)=-\frac{1}{4}\frac{\partial}{\partial\xi}\int_{-1}^1\mathrm{d}x\, x H^{q}_3(x,\xi,t),
  \\
  % G4
  \int_{-1}^1\mathrm{d}x\, x G^{q}_4(x,\xi,t)  &= -\frac{1}{4}\mathcal{G}^{q}_6(t)=-\frac{1}{4}\frac{\partial}{\partial\xi}\int_{-1}^1\mathrm{d}x\, x H^{q}_4(x,\xi,t),
   \\
  % G5
  \int_{-1}^1\mathrm{d}x\, x G^{q}_5(x,\xi,t)  &= 0,
  \\
  % G6
  \int_{-1}^1\mathrm{d}x\, x G^{q}_6(x,\xi,t) 
  &=-\frac{1}{2}\left[\mathcal{G}^{q}_5(t)+ \mathcal{G}^{q}_{10}(t)\right]=-\frac{1}{2}\int_{-1}^1\mathrm{d}x\, x H^{q}_2(x,\xi,t)+\frac{1}{2}\left[\tilde G_1(t)-\frac{t}{M^2}\tilde G_2(t)\right],
  \\
  % G7
  \int_{-1}^1\mathrm{d}x\, x G^{q}_7(x,\xi,t) 
  &=-\frac{\xi}{2}\mathcal{G}^{q}_6(t)=-\frac{1}{2}\int_{-1}^1\mathrm{d}x\, x H^{q}_4(x,\xi,t),
  \\
  % G8
  \int_{-1}^1\mathrm{d}x\, xG^{q}_8(x,\xi,t) 
  &=0,
  \\
  % G9
  \int_{-1}^1\mathrm{d}x\, x G^{q}_9(x,\xi,t)  &= 0,
\end{align}
\end{subequations}
where we have used Eq.~\eqref{G11}.

Based on Eq.~\eqref{Jq}, we find that the total quark or gluon AM in a state with maximal vector polarization along the $z$-direction can be expressed in terms of twist-2 GPDs as follows
\begin{equation}
J^z_a=\int_{-1}^1\mathrm{d}x\, \frac{x}{2}\left[H^{a}_2(x,0,0)+H^{a}_5(x,0,0)\right],
\end{equation}
which is nothing but Ji's relation~\cite{Ji:1996ek} for spin-1 targets. Summing over quark and gluon contributions, we recover the spin sum rule derived in~\cite{Abidin:2008ku}
\begin{equation}
J^z=\sum_{a=q,g}\int_{-1}^1\mathrm{d}x\, \frac{x}{2}\,H^{a}_2(x,0,0)=1.
\end{equation}
Unlike the case of spin-1/2 targets~\cite{Penttinen:2000dg,Kiptily:2002nx,Hatta:2012cs}, the quark OAM~\eqref{Lq} requires not only a pure twist-3 GPD but also a twist-2 GPD
\begin{equation}
L^z_q=\int_{-1}^1\mathrm{d}x\, x\left[\frac{1}{2}H^{q}_5(x,0,0)-G^{q}_6(x,0,0)\right].
\end{equation}
This twist-2 GPD contribution is associated with the tensor polarization and is therefore absent in the case of spin-1/2 targets.

Quark and gluon contributions to mass and pressure-volume work involve trace terms and hence twist-4 GPDs. Only the partial pressure-volume work anisotropy can be related to a twist-2 GPD
\begin{equation}
W^{ij}_a=-\mathcal T^{ij}M\int_{-1}^1\mathrm{d}x\, xH^{a}_5(x,0,0).
\end{equation}
Summing over quarks and gluons, the mass sum rule and the balance equations imply the following constraints
\begin{align}
    \sum_{a=q,g}\int_{-1}^1\mathrm{d}x\, xH^{a}_1(x,0,0)&=1 \label{eqn:H1:sum},\\
    \sum_{a=q,g}\int_{-1}^1\mathrm{d}x\, xH^{a}_5(x,0,0)&=0 \label{eqn:H5:sum}.
\end{align}
These relations can be interpreted as statements of energy-momentum
conservation for collinear parton distribution functions (PDFs).
Using the notation of \cite{Bacchetta:2000jk,Boer:2016xqr}
\footnote{Note that ref.~\cite{Boer:2016xqr} uses $(Pn)=1$.},
the unpolarized and tensor-polarized PDFs are given respectively by
$f_1^a(x) = H_1^a(x,0,0)$ and $f^a_{1LL}(x) = H^a_5(x,0,0)$.
In the case of quarks, they enter the deep inelastic structure functions $F_1$ and $b_1$ at leading order and leading twist as
\begin{align}
  F_1(x,Q^2) &=
  \frac{1}{2}\sum_{q} e_q^2  \left[
    H_1^q(x,0,0;\mu^2=Q^2) - H_1^q(-x,0,0;\mu^2=Q^2)
    \right],
  \\
  b_1(x,Q^2) &=
 \frac{1}{2} \sum_{q} e_q^2 \left[
    H_5^q(x,0,0;\mu^2=Q^2) - H_5^q(-x,0,0;\mu^2=Q^2)
    \right]
  .
\end{align}
The corresponding gluon PDFs mix with the quark ones  and contribute to the DIS structure functions  at higher order in the $\alpha_s$ perturbative expansion.
Thus Eq.~(\ref{eqn:H1:sum}) is a statement of the momentum sum rule
for PDFs.
The collinear structure functions for a scattering off a tensor polarized targets were first introduced in~\cite{Hoodbhoy:1988am,Jaffe:1989xy,Artru:1989zv} and the separate contributions of quarks and gluons to Eq.~(\ref{eqn:H5:sum}) were  previously
discussed in \cite{Efremov:1981vs,Efremov:1994xf}.

\section{Conclusion}
\label{sec:concl}
In this work, we found the most general form that the asymmetric,
gauge-invariant kinetic energy-momentum tensor (EMT) of a spin-$1$ hadron can take.
Expressions were given for both the full EMT
and the partial EMT due to a single parton type.
We explored the physical meaning of the gravitational form factors appearing in this EMT,
including sum rules imposed by conservation of momentum and angular momentum,
the decomposition of spin-$1$ hadron mass, and multipole moments of the EMT.
We also explored connections between the gravitational form factors
and other functions describing partonic structure,
such as axial form factors and generalized parton distributions up to twist three.

The spin-1 EMT was found to contain many more gravitational form factors than
the corresponding spin-$0$ or spin-$1/2$ EMTs.
A total of 11 form factors are present in the EMT decomposition,
with 9 of these in the symmetric part and 2 in the antisymmetric part. Among them, 6 structures have no analogues in the lower-spin cases and are related to the presence of tensor polarization modes. They
contribute to features new to spin-1 hadrons such as
a quadrupole moment
and possibly a non-zero intrinsic energy dipole moment aside from
the intrinsic angular momentum.

The structure of the spin-$1$ EMT is rich, and there remains much to be explored.
The pressure and shear force distributions encoded within it,
and how these differ from the simpler spin-$0$ case,
are worthy of detailed study.
It is also worth investigating how the EMT of a composite spin-$1$ hadron
compares to that of an elementary spin-$1$ particle,
such as a photon or one of the heavy electroweak gauge bosons.
These topics will be the subject of future work,
along with illustrative model calculations.

Experimentally, measurements of coherent hard exclusive processes with deuteron targets are possible at JLab and the future EIC with forward detectors.  Extraction of the chiral-even vector GPDs from these measurements would then constrain the deuteron gravitational form factors through the Mellin moments of these GPDs. Similarly, extraction of GDAs for the rho-rho meson pair from the crossed reaction $\gamma^*\gamma\to\rho\rho$ at Belle would constrain the rho meson gravitational form factors.

\subsection*{Note}
Shortly after the present work was completed and made available, the independent work~\cite{Polyakov:2019lbq}  appeared, dealing with the EMT of spin-1 hadrons. The results are consistent with ours and the document contains an especially useful comparison between the different nomenclature for the GFFs present in the literature.

\section*{Acknowledgments}

We are grateful to Bernard Pire and Maxim Polyakov for useful discussions. The work of SC and CL was supported by the Agence Nationale de la Recherche under the projects No.\ ANR-18-ERC1-0002 and ANR-16-CE31-0019.
AF was supported by the U.S.\ Department of Energy, Office of Science,
Office of Nuclear Physics, contract No.\ DE-AC02-06CH11357
and an LDRD initiative at Argonne National Laboratory
under Project No.\ 2017-058-N0.

\appendix
\section{Form factor counting}\label{AppA}

Let us consider the total number of GFFs we may
expect to appear in the decomposition of the spin-1 EMT.
This will depend on the EMT definition we use. First, let us consider the most general possible rank-2 tensor
that is even under charge conjugation.
It decomposes into several representations of the Lorentz group, namely $(0,0)$, $(1,0)\oplus(0,1)$, and $(1,1)$, which are respectively 1-, 6-, and 9-dimensional representations.
These representations have $J^{PC}$ quantum number decompositions
given in Table~\ref{tab:JPC:reps}.
\begin{table}[h]
  \begin{tabular}{c|c}
    Rep. & $J^{PC}$ \\ \hline
    $(0,0)$ & $0^{++}$ \\
    $(1,0)\oplus(0,1)$ & $1^{++}$, $1^{-+}$ \\
    $(1,1)$ & $0^{++}$, $1^{-+}$, $2^{++}$
  \end{tabular}
  \caption{
    $J^{PC}$ quantum numbers of several representations
    of the Lorentz group, constrained to be C-even.
  }
  \label{tab:JPC:reps}
\end{table}
With these $J^{PC}$ numbers,
we can use the method of Ji \& Lebed to count form factors~\cite{Ji:2000id}.

We consider a crossed channel matrix element
$\langle h\bar{h}|\hat{\mathcal{O}}|0\rangle$,
and write out all the possible $J^{PC}(L)$ quantum numbers that
$h\bar{h}$ can have up to $J=2$.
We then count the matches between these $J^{PC}$ and those
in our hypothetical rank-2 tensor.
Such lists can be found in Table~\ref{tab:JPC:LS}.
\begin{table}[h]
  \begin{tabular}{c|c}
    Spin & $J^{PC}_L(S)$ \\[1ex] \hline 
    0 & $0^{++}_0(0)$, $2^{++}_2(0)$ \\[1ex]
    $\frac{1}{2}$ & $0^{++}_1(1)$, $1^{++}_1(1)$,
      $2^{++}_1(1)$, $2^{++}_3(1)$ \\[1ex]
    1 & $0^{++}_0(0)$, $0^{++}_2(2)$,
      $1^{++}_2(2)$, $1^{-+}_1(1)$,
      $2^{++}_2(0)$, $2^{++}_{0,2,4}(2)$
  \end{tabular}
  \caption{
    Limited list of $J^{PC}_L(S)$ quantum numbers possible
    for a $h\bar{h}$ state.
    Only $J^{PC}$ numbers present in Table~\ref{tab:JPC:reps}
    are included here.
    The lists for spin-0 and spin-1 are from \cite{Liuti:2012INTTalk,Liuti:2019srv,Cosyn:2018thq},
    while the spin-1/2 list is from \cite{Ji:2000id}.
  }
  \label{tab:JPC:LS}
\end{table}
With these lists in hand, we can count the number of form factors
that are contributing to the general rank-2 tensor by each of the
representations it decomposes into.
The numbers are given in Table~\ref{tab:GFF:numbers}.
\begin{table}[h]
  \begin{tabular}{c|ccc}
    Rep. & ~Spin-0 GFFs~ & ~Spin-$\frac{1}{2}$ GFFs~ & ~Spin-1 GFFs~ \\[1ex] \hline
    $(0,0)$ & 1 & 1 & 2 \\
    $(1,0)\oplus(0,1)$ & 0 & 1 & 2 \\
    $(1,1)$ & 2 & 3 & 7
  \end{tabular}
  \caption{
    The number of form factors in a decomposition of a rank-2
    tensor belonging to any of the representations
    listed in the table.
  }
  \label{tab:GFF:numbers}
\end{table}

Each of the rows in Table~\ref{tab:GFF:numbers}
tells us something meaningful about form factor counts.
\begin{itemize}
  \item
    The first row gives us trace terms,
    which can be thought of as ``non-conserving'' terms.
    There is one such term for both spin-0 and spin-1/2,
    and they are associated with the $\bar{C}(t)$ GFFs.
    For spin-1, there are {\sl two} non-conserving terms of this kind.
  \item
    The second row tells us the number of GFFs appearing in the antisymmetric part of the EMT.
  \item
    The third row tells us the number of GFFs in the
    decomposition of a traceless symmetric rank-2 tensor,
    and they correspond to the number of GFFs that are known
    to appear in the second Mellin moments of twist-2 vector GPDs.
\end{itemize}

%%%%%%%%%%%%%%%%%%%%%%%%%%%%%%%%%%%%%%%%%%%%%%%%%%%%%%%%%%%%%%%%%%%%%%%%%%%%%%%%

\section{Parametrization with Lorentz projectors}\label{AppB}

The matrix elements of the most general local, gauge-invariant EMT for a massive spin-1 target can be written as
\begin{equation}
\langle p^\prime, \lambda^\prime \mid T_{\mu\nu}(0) \mid p, \lambda \rangle=\epsilon^{*\alpha}(p^\prime,\lambda^\prime)T_{\mu\nu,\alpha\beta}(P,\Delta)\epsilon^\beta(p,\lambda),
\end{equation}
where the effective vertex $T_{\mu\nu,\alpha\beta}(P,\Delta)$ is a Lorentz tensor constructed out of the invariant tensors $g_{\mu\nu}$ and $\epsilon_{\mu\nu\alpha\beta}$, and the available four-vectors $P$ and $\Delta$. Because of the onshell relations $(p^\prime\epsilon^{\prime*})=(p\epsilon)=0$, we choose to discard terms involving $P_\alpha$ or $P_\beta$ as they are not independent from those involving $\Delta_\alpha$ or $\Delta_\beta$. Finally, discrete symmetries impose further constraints on the effective vertex. Invariance under parity and time reversal implies that
\begin{equation}
T_{\mu\nu,\alpha\beta}(P,\Delta)=T_{\bar\mu\bar\nu,\bar\alpha\bar\beta}(\bar P,\bar\Delta)=T^*_{\bar\mu\bar\nu,\bar\alpha\bar\beta}(\bar P,\bar\Delta),
\end{equation}
where a bar over a four-vector or a Lorentz index stands for the parity-transformed object $\bar a^\mu=a^{\bar\mu}=(a^0,-\boldsymbol a)$. The hermiticity property leads to the last constraint
\begin{equation}
T_{\mu\nu,\alpha\beta}(P,\Delta)=T^*_{\mu\nu,\beta\alpha}(P,-\Delta).
\end{equation}
It follows from these constraints that the effective vertex is real, and that its symmetric part under the exchange $\alpha\leftrightarrow\beta$ involves only even powers of $\Delta$, whereas the antisymmetric part involves only odd powers of $\Delta$. Note also that the Levi-Civita tensor need not be considered since it is the only object with negative intrinsic parity and since the product $\epsilon_{\mu\nu\alpha\beta}\epsilon_{\rho\sigma\tau\lambda}$ can be rewritten in terms of the metric only.

Interestingly, we can find a complete parametrization of the effective vertex in terms of the Lorentz projectors onto $(0,0)$, $(1,0)\oplus(0,1)$, and $(1,1)$ representations
\begin{subequations}
\begin{align}
(I_{\mu\nu})_{\alpha\beta}&=\frac{1}{4}g_{\mu\nu}g_{\alpha\beta},\\
(D_{\mu\nu})_{\alpha\beta}&=\frac{1}{2}(g_{\mu\alpha}g_{\nu\beta}-g_{\mu\beta}g_{\nu\alpha}),\\
(Q_{\mu\nu})_{\alpha\beta}&=\frac{1}{2}(g_{\mu\alpha}g_{\nu\beta}+g_{\mu\beta}g_{\nu\alpha})-\frac{1}{4}g_{\mu\nu}g_{\alpha\beta}.
\end{align}
\end{subequations}
We recognize in particular the generators of Lorentz transformations $(M^{\mu\nu})^{\alpha\beta}=2i(D^{\mu\nu})^{\alpha\beta}$ associated with the vector representation $(\frac{1}{2},\frac{1}{2})$. Taking into account all the constraints discussed above, we find that the most general effective vertex can be written as a linear combination of the following eleven independent Lorentz structures (omitting the $\alpha,\beta$ indices for convenience)
\begin{equation}\label{covcount}
\begin{split}
&I_{\mu\nu},P_\mu I_{\nu P},\Delta_\mu I_{\nu \Delta},\\
&P_\mu D_{\nu \Delta},P_\nu D_{\mu \Delta},\\
&Q_{\mu\nu},\Delta_\mu Q_{\nu\Delta},\Delta_\nu Q_{\mu\Delta},g_{\mu\nu}Q_{\Delta\Delta},P_\mu P_\nu Q_{\Delta\Delta},\Delta_\mu\Delta_\nu Q_{\Delta\Delta}.
\end{split}
\end{equation}
At first sight, it seems that these are at odds with the counting in Appendix~\ref{AppA}, since we found three structures involving the projector onto the $(0,0)$ representation and six involving the projector onto the $(1,1)$ representation. In fact, the counting in Appendix~\ref{AppA} corresponds to the projections applied on the pair of indices $\mu\nu$. The structures corresponding to the $(0,0)$ part of the effective vertex are then the only two pure trace terms $I_{\mu\nu}$ and $g_{\mu\nu}Q_{\Delta\Delta}$. The structures corresponding to the $(1,1)$ part of the effective vertex are those obtained from removing the trace of the seven other symmetric terms  $P_\mu I_{\nu P}$, $\Delta_\mu I_{\nu \Delta}$, $Q_{\mu\nu}$, $\Delta_\mu Q_{\nu\Delta}$, $\Delta_\nu Q_{\mu\Delta}$, $P_\mu P_\nu Q_{\Delta\Delta}$, and $\Delta_\mu\Delta_\nu Q_{\Delta\Delta}$.

This technique has the advantage of showing the connection with lower spin targets. For spin-$0$ targets we have $I_{\mu\nu}=\frac{1}{4} g_{\mu\nu}$, $D_{\mu\nu}=0$ and $Q_{\mu\nu}=0$, and for spin-$1/2$ targets we have $I_{\mu\nu}=\frac{1}{4} g_{\mu\nu}\mathbb I$, $D_{\mu\nu}=i\sigma_{\mu\nu}$ and $Q_{\mu\nu}=0$\footnote{Interestingly, since $\sigma^{\mu\nu}=\frac{i}{2}[\gamma^\mu,\gamma^\nu]$ we can interpret the Clifford algebra $\{\gamma^\mu,\gamma^\nu\}=2g^{\mu\nu}\mathbb I$ as the condition of vanishing quadrupole.}. According to~\eqref{covcount} this leads respectively to 3 and 5 GFFs, in agreement with Appendix~\ref{AppA}.

\section{Polarization four-vector bilinears}\label{AppC}

Matrix elements of spin-1 targets can be written in terms of polarization four-vector bilinears contracted with some Lorentz tensors
\begin{equation}
\langle p^\prime, \lambda^\prime \mid O^{\mu_1\cdots\mu_n}(0) \mid p, \lambda \rangle=\epsilon^*_\alpha(p^\prime,\lambda^\prime)O^{\mu_1\cdots\mu_n,\alpha\beta}(P,\Delta)\epsilon_\beta(p,\lambda).
\end{equation}
In the forward limit, the polarization four-vector bilinear reduces to the covariant density matrix which can be written as~\cite{Joos:1964,Zwanziger:1965}
\begin{equation}
\epsilon_\beta(p,\lambda)\epsilon^*_\alpha(p,\lambda^\prime)=\rho_{\beta\alpha}(p)=\left[\frac{1}{3}\mathbb I+\frac{1}{2}\mathcal S^\mu(p)\Sigma_\mu+\mathcal T^{\mu\nu}(p)\Sigma_{\mu\nu}\right]_{\beta\alpha},
\end{equation}
where the vector and tensor polarization operators read
\begin{align}
\Sigma_\mu&=\frac{1}{2M}\epsilon_{\mu\rho\sigma p}M^{\rho\sigma},\\
\Sigma_{\mu\nu}&=-\Sigma_{\{\mu}\Sigma_{\nu\}}+\frac{1}{3}\left(g_{\mu\nu}-\frac{p_\mu p_\nu}{M^2}\right)\Sigma\cdot\Sigma,
\end{align}
and are orthogonal to $p^\mu$ as required by the onshell relation $(p\epsilon)=0$. Clearly, the covariant vector and tensor polarizations can be taken so as to satisfy the same properties as the associated operators
\begin{align}
p_\mu\mathcal S^\mu(p)&=0,\\
\mathcal T^{\nu\mu}(p)&= \mathcal T^{\mu\nu}(p),\\
p_\mu\mathcal T^{\mu\nu}(p)&=0,\\
g_{\mu\nu}\mathcal T^{\mu\nu}(p)&=0.
\end{align}
They carry the dependence on the polarizations $\lambda$ and $\lambda'$, which has been omitted for convenience.

The generators of Lorentz transformations in the $(\frac{1}{2},\frac{1}{2})$ representation being given by
\begin{equation}
(M^{\rho\sigma})_{\alpha\beta}=i(\delta^\rho_\alpha\delta^\sigma_\beta-\delta^\rho_\beta\delta^\sigma_\alpha),
\end{equation}
the unit, vector and tensor polarization operators take the explicit form
\begin{align}
(\mathbb I)_{\beta\alpha}&=-P_{\beta\alpha},\\
(\Sigma_\mu)_{\beta\alpha}&=\frac{i}{M}\epsilon_{\mu\beta\alpha p},\\
(\Sigma_{\mu\nu})_{\beta\alpha}&=-P_{\mu\{\beta}P_{\alpha\}\nu}+\frac{1}{3}P_{\mu\nu}P_{\beta\alpha},
\end{align}
where $P_{\mu\nu}$ is the projector onto the subspace orthogonal to $p^\mu$ given by
\begin{equation}
P_{\mu\nu}=g_{\mu\nu}-\frac{p_\mu p_\nu}{M^2}.
\end{equation}
All these operators are orthogonal to each other, so that the normalization, covariant vector and tensor polarizations can simply be obtained as follows
\begin{align}
\textrm{Tr}[\rho(p)\mathbb I]&=1,\\
\textrm{Tr}[\rho(p)\Sigma^\mu]&=-\mathcal S^\mu(p),\\
\textrm{Tr}[\rho(p)\Sigma^{\mu\nu}]&=\mathcal T^{\mu\nu}(p).
\end{align}

Denoting the rest-frame four-momentum as $k^\mu=Mg^{\mu 0}$, the general polarization four-vector bilinear can be written as
\begin{equation}
\epsilon^\beta(p,\lambda)\epsilon^{*\alpha}(p^\prime,\lambda^\prime)=(\Lambda_\text{can})^\beta_{\phantom{\beta}\rho}\,(\Lambda^\prime_\text{can})^\alpha_{\phantom{\alpha}\sigma}\,\rho^{\rho\sigma}(k),
\end{equation}
where the standard canonical rotationless Lorentz boost tensor reads
\begin{equation}
(\Lambda_\text{can})^\mu_{\phantom{\mu}\nu}=\begin{pmatrix}
\frac{p^0}{M}&\frac{p^1}{M}&\frac{p^2}{M}&\frac{p^3}{M}\\
\frac{p^1}{M}&1+\frac{(p^1)^2}{M(p^0+M)}&\frac{p^1p^2}{M(p^0+M)}&\frac{p^1p^3}{M(p^0+M)}\\
\frac{p^2}{M}&\frac{p^1p^2}{M(p^0+M)}&1+\frac{(p^2)^2}{M(p^0+M)}&\frac{p^2p^3}{M(p^0+M)}\\
\frac{p^3}{M}&\frac{p^1p^3}{M(p^0+M)}&\frac{p^2p^3}{M(p^0+M)}&1+\frac{(p^3)^2}{M(p^0+M)},
\end{pmatrix}
\end{equation}
or in a more covariant form~\cite{Krause:1977yr,doi:10.1119/1.12748}
\begin{equation}
(\Lambda_\text{can})^\mu_{\phantom{\mu}\nu}=\delta^\mu_\nu-\frac{(p+k)^\mu(p+k)_\nu}{k\cdot(p+k)}+\frac{2p^\mu k_\nu}{M^2}.
\end{equation}
Denoting for convenience the rest-frame covariant vector and tensor polarizations as
\begin{equation}
\mathcal S^\mu\equiv\mathcal S^\mu(k),\qquad \mathcal T^{\mu\nu}\equiv\mathcal T^{\mu\nu}(k),
\end{equation}
we find that the general polarization four-vector bilinear can be expressed as
\begin{equation}
\begin{aligned}
\epsilon^\beta(p,\lambda)\epsilon^{*\alpha}(p^\prime,\lambda^\prime)&=\frac{1}{3}\left[-g^{\beta\alpha}+\frac{p^\beta p^{\prime\alpha}}{M^2}-\frac{(p+k)^\beta\Delta^\alpha}{k\cdot(p+k)}+\frac{\Delta^\beta(p^\prime+k)^\alpha}{k\cdot(p^\prime+k)}+\frac{\Delta^2}{2}\frac{(p+k)^\beta(p^\prime+k)^\alpha}{[k\cdot(p+k)][k\cdot(p^\prime+k)]}\right]\\
&+\frac{i}{2M}\left[\epsilon^{\beta\alpha\mathcal Sk}-\epsilon^{\beta p^\prime\mathcal Sk}\frac{(p^\prime+k)^\alpha}{k\cdot(p^\prime+k)}-\frac{(p+k)^\beta}{k\cdot(p+k)}\epsilon^{p\alpha\mathcal Sk}+\epsilon^{pp^\prime\mathcal Sk}\frac{(p+k)^\beta(p^\prime+k)^\alpha}{[k\cdot(p+k)][k\cdot(p^\prime+k)]}\right]\\
&-\mathcal T^{\beta\alpha}+\mathcal T^{\beta p^\prime}\frac{(p^\prime+k)^\alpha}{k\cdot(p^\prime+k)}+\frac{(p+k)^\beta}{k\cdot(p+k)}\mathcal T^{p\alpha}-\mathcal T^{pp^\prime}\frac{(p+k)^\beta(p^\prime+k)^\alpha}{[k\cdot(p+k)][k\cdot(p^\prime+k)]}.
\end{aligned}
\end{equation}
In the forward limit, we recover the standard expression for the covariant density matrix
\begin{equation}
\rho_{\beta\alpha}(p)=-\frac{1}{3}P_{\beta\alpha}+\frac{i}{2M}\epsilon_{\beta\alpha\rho\sigma}\mathcal S^\rho(p)p^\sigma-\mathcal T_{\beta\alpha}(p),
\end{equation}
where
\begin{equation}
\mathcal S^\mu(p)=(\Lambda_\text{can})^\mu_{\phantom{\mu}\rho}\mathcal S^\rho,\qquad \mathcal T^{\mu\nu}(p)=(\Lambda_\text{can})^\mu_{\phantom{\mu}\rho}(\Lambda_\text{can})^\nu_{\phantom{\nu}\sigma}\mathcal T^{\rho\sigma}.
\end{equation}
As noticed in the case of Dirac bilinears~\cite{Lorce:2017isp}, in the forward limit $\Delta=0$ the dependence on the rest-frame four-momentum $k^\mu$ can be absorbed into the $p$-dependent covariant polarization. However, in the general off-forward case $\Delta\neq 0$ this is not possible anymore and the $k$-dependence remains explicit. This dependence is unavoidable in a relativistic theory and comes from the fact that, because of the non-commutativity of boosts, canonical polarization of a massive particle has to be defined from a boost of the rest-frame polarization to the frame of interest~\cite{Polyzou:2012ut}.

In the Breit frame $\vec P=\vec 0$, we find that the general polarization four-vector bilinear reduces to
\begin{equation}
\begin{aligned}
\epsilon^\beta(p,\lambda)\epsilon^{*\alpha}(p^\prime,\lambda^\prime)&=\frac{1}{3}\left[-g^{\beta\alpha}+\left(1+\frac{\Delta^2}{4M^2}\right)\frac{k^\beta k^\alpha}{M^2}-\frac{k^{[\beta}\Delta^{\alpha]}}{M^2}+\frac{\Delta^\beta\Delta^\alpha}{4M^2}\right]\\
&+\frac{i}{2M}\left[\epsilon^{\beta\alpha\mathcal Sk}-\frac{k^{\{\beta}\epsilon^{\alpha\}\Delta\mathcal S k}}{M^2}+\frac{\Delta^{[\beta}\epsilon^{\alpha]\Delta\mathcal S k}}{4M^2}\right]\\
&-\mathcal T^{\beta\alpha}-\frac{k^{[\beta}\mathcal T^{\alpha]\Delta}}{M^2}+\frac{\Delta^{\{\beta}\mathcal T^{\alpha\}\Delta}}{4M^2}+\frac{\mathcal T^{\Delta\Delta}}{4M^2}\frac{k^\beta k^\alpha}{M^2}+\mathcal O(\Delta^3).
\end{aligned}
\end{equation}

\section{Covariant parametrization of Generalized Parton Distributions}\label{AppD}

Just like for local operators, hadronic matrix elements of non-local operators can be parametrized in a covariant form if one includes in the list of available four-vectors the direction of non-locality, namely the lightlike four-vector $n$ in the case of parton distributions like GPDs. The most general parametrization of the quark GPD correlator~\eqref{GPDcorrelatorq}, that respects the constraints imposed by hermiticity, parity and time-reversal, reads
\begin{equation}
 V_{\lambda\lambda^\prime}^{\mu} = 
    \frac{1}{(Pn)}\epsilon^*_\alpha(p^\prime,\lambda^\prime)
     \mathcal V^{\mu,\alpha\beta}(P,\Delta,n)\,\epsilon_\beta(p,\lambda)
\end{equation}
with
\begin{equation}
\begin{aligned}
     \mathcal V^{\mu,\alpha\beta}(P,\Delta,n)
    &= g^{\alpha\beta}\left[P^\mu  F_1+\Delta^\mu
     F_2+\frac{M^2n^\mu}{(Pn)}\,F_3
     \right]\\
    &+\frac{\Delta^{\alpha}\Delta^{\beta}}{M^2}
     \left[P^\mu  F_4+\Delta^\mu
     F_5+\frac{M^2n^\mu}{(Pn)}\,F_6
     \right]\\
     &+\frac{M^2n^{\alpha}n^{\beta}}{(Pn)^2}
     \left[P^\mu  F_7+\Delta^\mu
     F_8+\frac{M^2n^\mu}{(Pn)}\,F_9
     \right]\\
     &+\frac{n^{\{\alpha} \Delta^{\beta\}}}{(Pn)}
     \left[P^\mu  F_{10}+\Delta^\mu
     F_{11}+\frac{M^2n^\mu}{(Pn)}\,F_{12}
     \right]\\
     &+\frac{n^{[\alpha} \Delta^{\beta]}}{(Pn)}
     \left[P^\mu  F_{13}+\Delta^\mu
     F_{14}+\frac{M^2n^\mu}{(Pn)}\,F_{15}
     \right]\\
     &+g^{\mu \{\alpha}
     \left[\Delta^{\beta\}}
     F_{16}+\frac{M^2n^{\beta\}}}{(Pn)}\,F_{17}
     \right]\\
    &+g^{\mu [\alpha}
     \left[\Delta^{\beta]}
     F_{18}+\frac{M^2n^{\beta]}}{(Pn)}\,F_{19}
     \right] ,
    \end{aligned}
\end{equation}
and  $F_i = F_i(x,\xi,t)$. Since $\mathcal V^{\mu,\alpha\beta}(P,\Delta,n)$ must be invariant under a rescaling of the lightlike direction $n\mapsto \alpha n$, factors of $M/(Pn)$ appear whenever necessary. Additional factors of the hadron mass $M$ have also been included to keep GPDs dimensionless. The relation between $F_i$ and the standard GPD basis $H_i,G_i$ of~\eqref{eq:vectorcorr:twist3} can easily be obtained by projection onto the twist-2 and 3 parts.

\bibliography{main}

\end{document}